**Correlation of size and oxygen bonding at the interface of Si nanocrystal in Si-SiO$_2$ nanocomposite: A Raman mapping study**

Ekta Rani[1], Alka A. Ingale[1], A. Chaturvedi[2], C. Kamal[3], D. M. Phase[4], M.  P. Joshi[2], A. Chakrabarti[3], A. Banerjee[5] and L. M. Kukreja[2]

[1]Laser Physics Applications Section, Raja Ramanna Centre For advanced Technology, Indore-452013, INDIA

[2]Laser Material Processing Division, Raja Ramanna Centre For advanced Technology, Indore-452013, INDIA

[3]Indus Synchrotrons Utilization Division, Raja Ramanna Centre For advanced Technology, Indore-452013, INDIA

[4]UGC-DAE Consortium for Scientific Research, Indore 452010, India

[5]BARC Training School at RRCAT, Raja Ramanna Centre For advanced Technology, Indore-452013, INDIA

E-mail: alka@rrcat.gov.in

**Abstract.** Si-SiO$_2$ multilayer nanocomposite (NCp) films, grown using pulsed laser deposition with varying Si deposition time are investigated using Raman spectroscopy/mapping for studying the variation of Si phonon frequency observed in these NCps.  The lower frequency (LF) phonons (~ 495 - 510 cm$^{-1}$) and higher frequency (HF) phonons (~ 515 - 519 cm$^{-1}$) observed in Raman mapping data (Fig. 1A) in all samples studied are attributed to have originated from surface (Si-SiO$_2$ interface) and core of Si nanocrystals, respectively. The consistent picture of this understanding is developed using Raman spectroscopy monitored laser heating/annealing and cooling (LHC) experiment at the site of a desired frequency chosen with the help of Raman mapping, which brings out clear difference between core and surface (interface) phonons of Si nanocrystals. In order to further support our attribution of LF being surface (interface) phonons, Raman spectra calculations for Si$_{41}$ cluster with oxygen termination are performed which shows strong Si phonon frequency at 512 cm$^{-1}$ corresponding to the surface Si atoms. This can be considered analogous to the observed phonon frequencies in the range 495 - 510 cm$^{-1}$ originating at the Si-SiO$_2$ interface (extended). These results along with XPS data show that nature of interface (oxygen bonding) in turn depends on the size of nanocrystals and thus LF phonons originate at the surface of smaller Si nanocrystals. The understanding developed can be extended to explain large variation observed in Si phonon frequencies of Si-SiO$_2$ nanocomposites reported in the literature, especially lower frequencies.

Keywords: Nanocomposite, Raman spectroscopy, Si nanocrystals, Si-SiO$_2$ interface, DFT/TDDFT calculations.

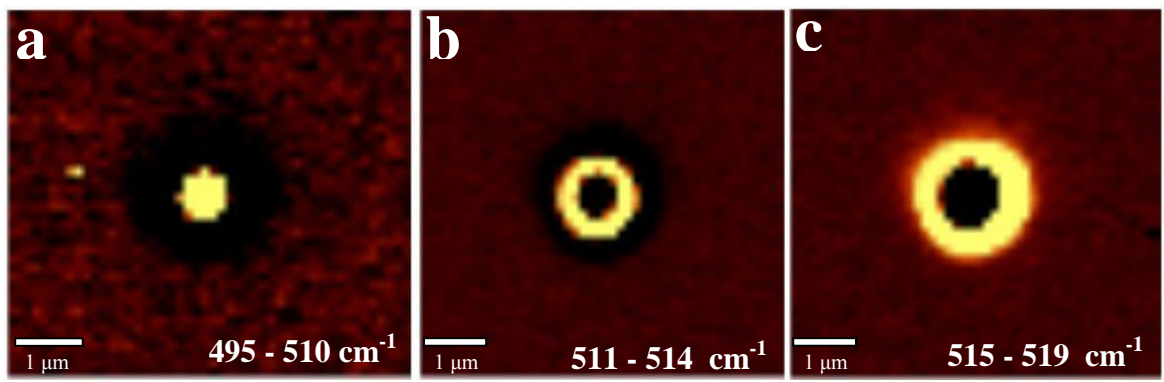

Fig. 1A. Representative Raman image of sample E1 generated using intensity of phonons with frequencies in the range a) 495 - 510 cm$^{-1}$, b) 511 - 514 cm$^{-1}$ and c) 515 - 519 cm$^{-1}$.

## 1. Introduction

Low dimensional Si nanostructures have been extensively studied because of their fundamental scientific interest, application in optoelectronic[1] and photovoltaic[2] devices. Most of the properties exhibited by nanocrystals (NCs) are size dependent due to quantum confinement effect and/or increase in surface to volume ratio. Due to large surface to volume ratio in NCs, two possibilities can arise, 1) Agglomeration of particles to form larger particles and 2) large number of surface defect states which can suppress their luminescence properties. This may wipe out benefits associated with NCs. Nanocomposite (NCp) of semiconductor nanostructures in a suitable inorganic/organic matrix is one of the ideal solutions to this problem, which have many additional advantages. Different types of insulating materials which have been used as matrix for Si NCs are $SiO_2$, $Al_2O_3$, $Si_3N_4$, and SiC.[3-6] Out of these, especially $SiO_2$ matrix has been used to embed Si NCs to grow multilayer structure since multilayer structure can overcome the Shockley-Queisser limit of 30% efficiency.[7] Raman mapping/spectroscopy is an ideal technique to study NCps, as it gives information about local bonding environment at the surface of NCs, which may change within a NCp due to presence of surrounding matrix. Noninvasive nature of Raman spectroscopy makes it more suitable local probe for understanding the nature of the NCp. Si (NCs or bulk) is one of the most studied materials using Raman spectroscopy, nevertheless repeatability and corroboration to other properties with Raman spectra has been difficult for different forms of Si (porous Si) or Si in different environment. Further, irrespective of growth technique used; variation in Si phonon frequency (in the range 495 - 519 $cm^{-1}$) in $Si$-$SiO_2$ nanocomposite is reported in the literature.[8-37] There have been few attempts to consolidate the findings.[9, 22, 25, 27] However, the variation of Si phonon frequency observed in these NCps cannot be completely explained using presently proposed models.[8, 22, 25, 38] In the range noted above, origin of higher frequency phonon ~ 516 cm$^{-1}$ has been attributed to phonon confinement in Si NCs and is well understood. However, understanding of origin of lower frequency phonons ~ 500 cm$^{-1}$ is far from satisfactory. Different groups have reported different Si phonon frequencies in the range noted above, using different excitations for Si-SiO$_2$ NCp grown with different methods and under different conditions. In most of the data reported, the understanding is given with respect to specific growth conditions. In addition, there have been some attempts to understand this variation reported in the literature, but no satisfactory understanding could be reached for low frequency phonons. The aim of this investigation is to address the origin of these low frequency phonons. This is the first time that the whole range of Si phonon frequencies are reported in one sample using Raman mapping. This allowed us to do systematic investigation of the same for 6 such Si-SiO$_2$ samples.

In order to understand the origin of this variation, we carry out laser heating/annealing & cooling (LHC) experiment at desired frequency. This variation of the Si phonon frequency below 510 cm$^{-1}$ NCps has been attributed due to the local variation (oxygen environment) at the surface of smaller Si NCs i.e. Si NCs-SiO$_2$ interface in the SiO$_2$ matrix. In the process of this analysis, consolidation and understanding of the data available in the literature is also achieved. To gain more insight into the matter, we carry out *ab initio* density functional theory (DFT) and time-dependent density functional theory (TDDFT) based calculations of Raman spectra of optimized Si cluster (Si$_{41}$) terminated by O and H atoms. Furthermore, we have performed X-ray photoelectron spectroscopy (XPS) measurements on these samples to study the extended interface.

The present paper is organized in the following manner. In section 2, Raman spectroscopy and mapping experiments are described. Section 3 and 4 are devoted to Raman spectroscopy and Raman mapping results showing spatial variation in Si phonon frequency. In section 5, we discuss these results to construct a consistent picture based on our understanding that low frequencies (in the range 495 - 510 cm$^{-1}$) and high frequencies (in the range 515 - 519 cm$^{-1}$) Si phonons originate from different parts of a Si nanocrystal in Si-SiO$_2$ nanocomposite. In the same section, the results of calculated Raman spectra on Si$_{41}$ cluster are also presented. The understanding developed from LHC experiment and it's corroboration with the calculated Raman spectra and XPS results are summarized in the conclusion.

## 2.    Experimental

Si-SiO$_2$ multilayer nanocomposites are grown using PLD. Silicon NCs of variable mean sizes are grown by varying the deposition time of Si from 45 to 210 s and SiO$_2$ layer deposition time is kept constant for all samples. The total thickness is expected to vary from 360 nm to 560 nm with the increase in deposition time from 45 s to 210 s. These multilayer (15 multilayer (Si/SiO$_2$)) films are subsequently annealed at 800 $^{\circ}$C in the PLD chamber in the nitrogen ambiance. For further growth details one is referred to our earlier paper.[39] We have studied 6 samples with varying Si deposition time and these samples are designated as 1) E1 (45s), 2) E2 (90 s), 3) E3 (120 s), 4) E4 (150 s), 5) E5 (180 s), and 6) E6 (210 s) with numbers appearing in the parenthesis denoting the deposition time in seconds. Micro-Raman spectroscopic/mapping measurements are performed in backscattering geometry at room temperature using Acton 2500i (single) monochromator with air cooled CCD detector, a part of scanning probe microscopy_integrated Raman system set up, WiTec (Germany). The Raman data is excited with 441.6 nm wavelength of He-Cd laser using a 50x microscope objective lens (numerical aperture 0.55) giving spatial resolution of ~ 0.98 μm. Raman spectrum is collected in the back scattering geometry and dispersed using single spectrometer of half meter focal length. Spectral resolution of the system used for the measurements is ~ 4 cm$^{-1}$. Notch filter is used for Rayleigh light rejection. Data is collected using air cooled charged coupled device (CCD) detector. XPS measurements are performed at room temperature with an OMICRON 180° hemispherical analyzer (model EA 125) using Al Kα (photon energy = 1486.7eV) source. The hemispherical analyzer is operated in constant-pass energy mode and its overall energy resolution with pass energy of 50 eV was estimated to be 0.8 eV. The measurements are carried out with a photoelectron take-off angle of 45° and the pressure in the spectrometer chamber during the measurements is ~ 10$^{-10}$ mbar. The beam size in XPS is ~ 1 cm * 1cm covering the full sample.

### 3. Raman spectroscopy

When measured at spatially different positions, Raman spectroscopic measurements of Si-SiO$_2$ NCp/films show variation in the Si phonon frequency from 495 to 519 cm$^{-1}$ in all the samples (numbered as E1-E6). It is well known that Raman spectra of Si NCs show asymmetric phonon line shape and red shift in phonon frequency due to phonon confinement effect accounted by phonon confinement model (PCM).[40, 41] Primarily PCM has been used to explain Raman spectroscopy results reported for Si NCs grown using various techniques.[8-9, 31, 42-44] Further, in the literature, it is already noted that PCM alone is inadequate to give correct red shift and line shape fitting simultaneously specially for frequencies below 510 cm$^{-1}$.[21-22, 25, 33, 44-46] Several groups have addressed this discrepancy and different solutions have been proposed to obtain correct red shift along with correct line shape fitting as briefly accounted below.

Bond Polarizability (BPL) model[45] has been used for calculating sizes (<50 Å) of Si NCs from the observed red shift[22, 24-25, 47] whereas; PCM is used for calculating their line shape. However, even BPL model is unable to give accurate size for Si phonon frequency below 512

cm$^{-1}$ [48] i.e. BPL model does not explain the occurrence of low phonon frequencies. Modified PCM accounting for size distribution can describe only the phonon frequencies in the range 516-522 cm$^{-1}$.[8, 38] Further, red shifts of ~ 2 cm$^{-1}$ of Si phonons due to laser heating had been reported for free Si NCs or embedded in SiO$_2$ grown on Si substrate at laser power density (PD): 5-10 kW/cm$^2$.[38, 49-50] Thus red shift of the order of ~ 25 cm$^{-1}$ as in our case, cannot be understood considering any of the above explanations.

## 4.  Raman mapping: Si phonon peak variation

Many sets of Raman mapping (acquisition time 1 s, step size ~ 0.15 μm) are performed on each of the six samples to generate enough statistical data to make appropriate observations. Since, Raman spectra show phonon frequencies in the range 495 - 519 cm$^{-1}$ for all the samples, hence Raman images are generated using intensity of phonon frequencies in the range 495 - 519 cm$^{-1}$. Representative Raman image for sample E1 is shown in Fig. 1(a). It is important to note that the spatial resolution of the Raman microscope is ~ 0.98 μm; however, since low laser intensity and low acquisition times are used to avoid laser heating effects, only central part of Gaussian beam profile with highest intensity gives Raman signal leading to better spatial resolution than expected.  The image shows signal from crystalline Si in well separated regions dispersed over the whole film. This suggests that although the growth is multilayer, evidence of continuous film of crystalline Si is not observed.  Instead, formation of crystalline Si is found only in certain areas of the film.  Fig. 1(b) shows corresponding Raman spectra for the marked positions in Fig. 1(a), showing presence of different Si phonon frequencies. There are two important observations to be noted, i) at the sites where these frequencies are measured, no significant amorphous content is observed, and ii)  presence of three different kind of line shapes for observed range of Si phonons. The higher frequency phonons above ~ 515 cm$^{-1}$ (Fig. 1(b)) show asymmetric line shape, which can be well fitted using PCM. For NCs, quantum confinement of phonons due to size leads to breakdown of Raman selection rule ($\mathbf{q} \sim 0$), leading to contribution from $\mathbf{q} \neq 0$ phonons to the Raman spectrum, where $\mathbf{q}$ is a wave vector of scattered phonon. This leads to red shift, broadening and asymmetric line shape of phonons. We find that asymmetric line shape of phonons in the range 515 - 519 cm$^{-1}$ can be well fitted using PCM given by Campbell and Fauchet.[40] According to this model, first order Raman spectrum of a nanoparticle can be calculated as,

$$I(\omega) \, \alpha \, \int \frac{|C(q)|^2}{(\omega - \omega(q))^2 + \left(\Gamma_0/2\right)^2} d^3 q$$

here, $\omega(\mathbf{q}) =$ is the phonon dispersion curve and $\Gamma_0$ is the natural line width of the zone-center optical phonon. Phonon dispersion curve that we have used has been determined by fitting neutron scattering data obtained for bulk Si[41]

$$\omega \, (\mathbf{q}) = \, \left(A + B \cos\left(\frac{\pi q}{2}\right)\right)^{1/2}$$

Where, $A = 1.714*10^5$ cm$^{-2}$ and $B = 1.00*10^5$ cm$^{-2}$. The fitting suggests that these phonons originate from Si NCs having sizes $\sim$ 70 - 100 Å.

It is interesting to note that lower phonon frequencies i.e. below $\sim$ 510 cm$^{-1}$ show Lorentzian line shape (Fig. 1(b)). It is important to note here that at very few places (5%), Raman signal corresponding to amorphous Si is observed, where these Si phonon frequencies are measured. In these cases, small asymmetry can be deconvoluted as due to this small amorphous content as broad small peak at $\sim$ 480 cm$^{-1}$ and a Lorentzian shape for a low frequency phonon. The amorphous content of Si nanoparticle is known to depend on growth method as well as growth parameters.[9, 17, 24, 25] It is further important to note that earlier, no emphasize is given to the Lorentzian line shape of low frequency phonons. This is probably because no Raman mapping has been reported so far for this NCps and therefore spatial variation in the Raman spectra in one sample has not been noted earlier in such details. We consider three different possibilities for the occurrence of these low frequency phonons. i) the line shape suggests that these phonons are coming from large size Si NCs (nearly bulk > 300A), ii) they originate at surface/interface of Si-SiO$_2$ nanocomposite and iii) they are originating at amorphous SiO$_2$ matrix. If they are coming from large NCs, then the frequency of the phonons is expected to be ~ 521 cm$^{-1}$. The lower frequency with large size NC can be explained if there is tensile strain in the Si NC. However, this possibility can be rejected as enough evidences are there in literature, which suggests growth of Si NCs in SiO$_2$ matrix leads to compressive stress in NCs[20, 38] leading to blue shift in the phonon frequency. Further, Raman spectra of a- SiO$_2$ shows broad peak at 800 cm$^{-1}$, therefore third possibility is also rejected. This leaves us with the possibility of these phonons originating from surface/interface of Si-SiO$_2$ nanocomposite. Further, we find that the asymmetric line shape of phonons in the intermediate frequency range 511-514 cm$^{-1}$ (Fig. 1(b)) cannot be fitted using PCM and hence this frequencies cannot be said to be originating from confinement due to small size Si NC. Based on these line shape differences, these phonons are broadly separated into three different frequency regions as follows, i) 495 - 510 cm$^{-1}$ ( low frequency - LF ) phonons with Lorentzian line shape, ii) 515 - 519 cm$^{-1}$ (high frequency - HF) phonons with asymmetric line shape and iii) 515 - 519 cm$^{-1}$ (intermediate frequency - IF) phonons for the convenience of further investigation. By selecting LF, IF and HF range phonons, separate Raman images of the selected area Raman mapping are created. Representative Raman images for LF, IF and HF phonons for the sample E1 are shown in Fig. 1(c)-1(e). Lower frequency (than that of bulk) Si phonons are observed only at some sites (not all over) clustered together in the films in Raman mapping indicating clustering of Si NCs in these nanocomposite films. This cluster formation is being further investigated further using Raman and AFM mapping and details will be published elsewhere.[51] For our studies here, it is important to know that although growth is multilayer the deposition process has given rise to clusters of Si NCs embedded in $SiO_2$ matrix and their density increases with increase in number of layers.[51]

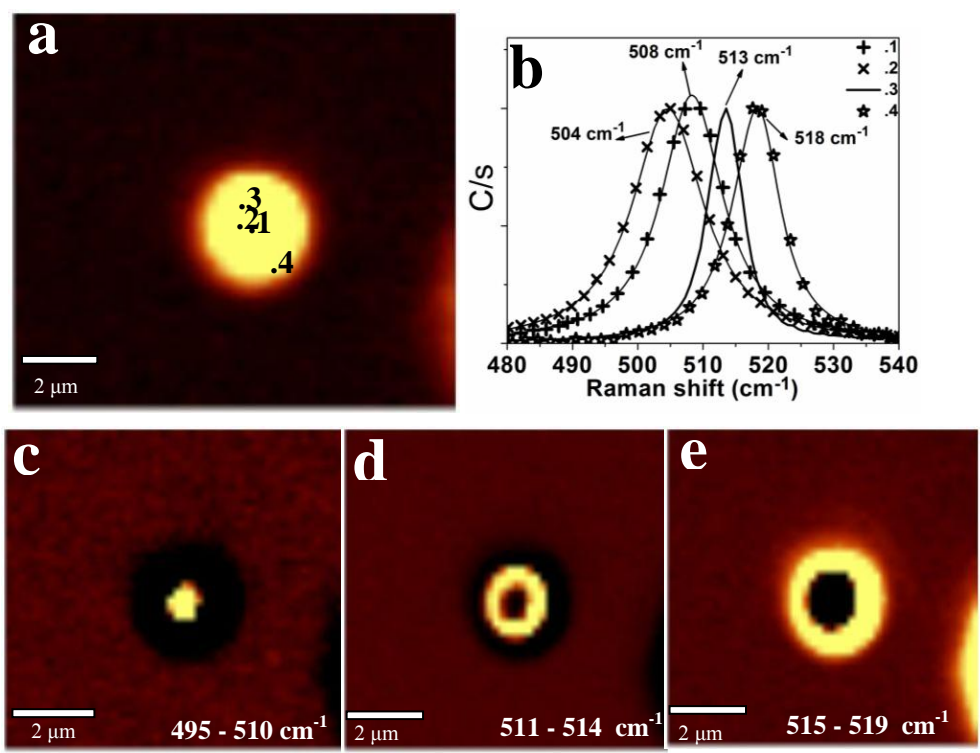

Fig. 1. a) Representative Raman image of sample E1 generated using frequency in the range 495 - 525 cm$^{-1}$ and (b) Corresponding Raman spectra on positions as marked in Raman image (a) showing Lorentzian line shape for phonon frequencies ~ 504 and 508 cm$^{-1}$, PCM fit asymmetric line shape for phonon frequency ~ 518 cm$^{-1}$ and asymmetric line shape for phonon frequency ~ 513 cm$^{-1}$. Raman image generated using phonon frequency c) ~ 495 - 510 cm$^{-1}$, d) 511 - 514 cm$^{-1}$, e) ~ 515 - 519 cm$^{-1}$.

Furthermore, while performing Raman mapping, we have also observed simultaneous occurrence of LF and HF phonons ~ 501 cm$^{-1}$ and ~ 517 cm$^{-1}$ respectively, at some spatial positions. Similar kind of observation is reported earlier by Faraci et al.[22] in Si-SiO$_2$ nanocomposite. They have suggested that lower frequencies may be due to Si aggregates having a thin SiO$_x$ (with x = 1 − 2) interface for Si NCs in SiO$_2$ matrix.[22]

We have performed laser heating/annealing and cooling (LHC) experiment on these four kinds of Si phonon peaks observed i) LF phonons, ii) IF phonons, iii) HF phonons and iv) simultaneously observed LF and HF phonons. This experiment is designed to know, if the origins of LF, IF and HF phonons are same or different as discussed above and if we can get further clue to their origins. It is also interesting to note here that low and high frequency range Si phonons behave markedly differently under LHC experiment justifying the broad categorization being made based on the line shape analysis.

*4.1. Raman spectroscopy monitored Laser heating/annealing and cooling (LHC) experiment:* Steps of LHC experiment are,

1. The desired peak frequency is located using Raman mapping for doing LHC experiment on the said frequency ranges.

2. Laser heating and cooling steps:

1$^{st}$ step; 2$^{nd}$ step : *laser on for 16 mins*; 3$^{rd}$ step : *Laser shutter off for 40 mins*; 4$^{th}$ step : *Again laser on for 16 mins* ; 5$^{th}$ step: *Laser shutter off for 3 hrs.*

At the end of each step (including 1$^{st}$ step: initiation of the experiment) Raman spectrum is recorded with acquisition time 3 s. LHC experiment is performed on each sample for at least 2-3 frequencies in each of the three phonon ranges (LF, IF, HF) noted earlier. Laser heating leading to redshift of Si phonon of Si NC is avoided by keeping low laser PD ~ 5 kW/cm$^2$. This is also further confirmed by the fact that all frequencies including LF are observed in one Raman mapping (Fig.1a & b) experiment at a particular laser PD. Therefore, one cannot attribute LF and IF phonons as due to laser heating. This further indicates that low frequency phonons are intrinsic to Si-SiO$_2$ nanocomposite. While performing this experiment, no mechanical movement is involved.

The effect of LHC experiment on LF, IF and HF phonons is found to be quite different as noted earlier. It is important to note that observations for LF and HF phonons during LHC are same irrespective of their simultaneous or separate occurrence in the Raman spectra. The results are summarized in the following.

i)   LF phonons (495 - 510 cm$^{-1}$) **:** In all the measurements performed in this frequency range, the behavior of peak frequencies is similar i.e. it blue shifts during the LHC experiment. Representative Raman spectra for phonon frequency ~ 499 cm$^{-1}$ during LHC experiment is shown in Fig. 2(b).  This peak shows continuous blue shift of ~ 10 cm$^{-1}$. Further, crystalline order has increased on laser heating (annealing) as FWHM has decreased by ~ 4.5 cm$^{-1}$ during the experiment.

ii)  HF phonons (515 - 519 cm$^{-1}$) **:**  In all the measurements performed in this frequency range, phonons show similar behavior i.e. it does not change during the LHC experiment. Representative Raman spectra for phonon frequency ~ 516 cm$^{-1}$ during LHC experiment is shown in Fig. 2(c).

iii) Simultaneously observed LF and HF phonons ~ 501 cm$^{-1}$ and ~ 517 cm$^{-1}$ respectively

iv) The two peaks Raman spectra (Fig. 2(a)) are approximated as two Lorentzian for convenience of noting the variation in the frequency. Out of the two peaks LF phonon shows total blue shift by ~ 4 cm⁻¹, whereas HF phonon does not show any significant change.

v) IF phonons: Phonons in this frequency range follows LF behavior during LHC experiment, however, shows some differences with respect to LF phonons. Details will be published elsewhere.[52]

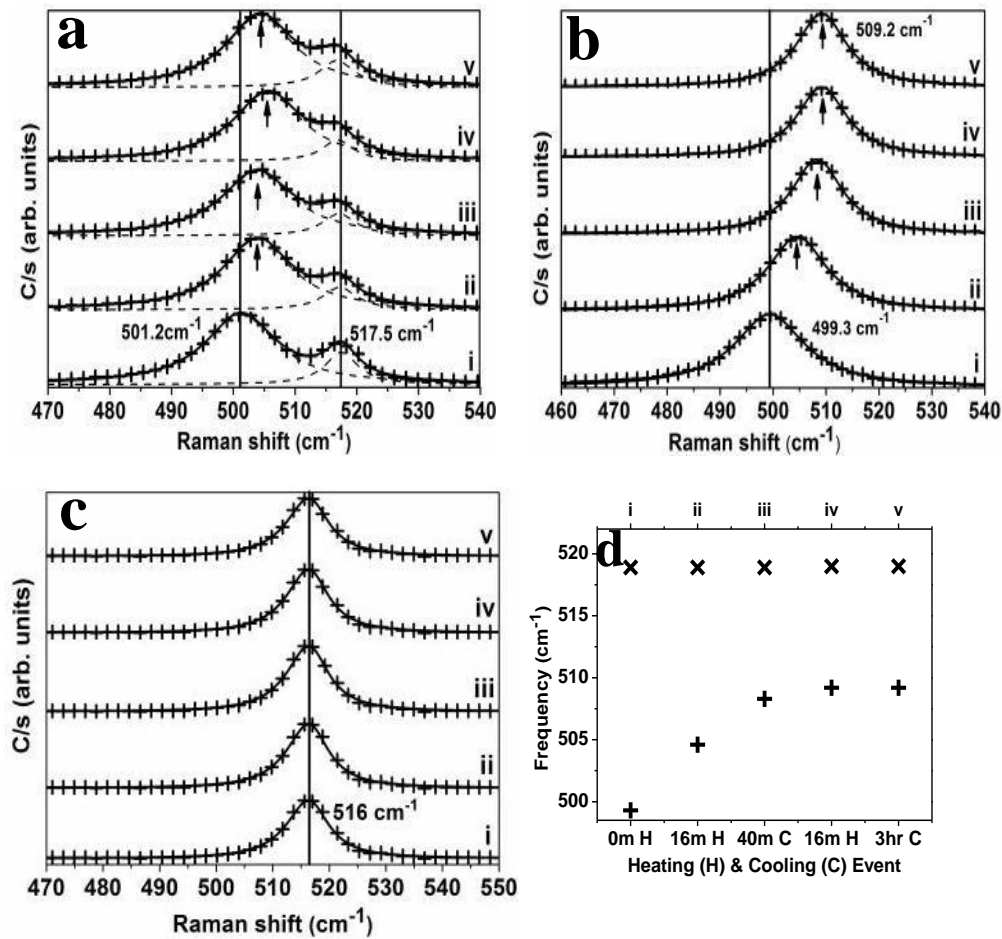

Fig. 2. Representative Raman spectra showing change for a) two resolvable phonon frequencies occurring simultaneously ~ 501 and 517 cm⁻¹, b) phonon frequency ~ 499 cm⁻¹, c) phonon frequency ~ 516 cm⁻¹ occuring separately during 5 steps of LHC experiment as (i) 0 min heating, (ii) 16 min heating, (iii) 40 min cooling, (iv) again 16 min heating and (v) 3hr cooling. Raw data points are shown by (+) and the solid line is the line shape fit to the spectra as noted in the text and d) Change in LF and HF phonon during LHC process. Symbols (+) and (×) show phonon frequency ~ 499 cm⁻¹ and 517 cm⁻¹.

The variation of peak frequency obtained in 5 steps of heating/annealing and cooling experiment for different range peak frequencies is summarized in Fig. 2(d).

## 5. Discussion

The differences noted in the behavior of LF and HF phonons clearly indicate that they have different origins. The manifold reasons as given below and consistencies in the observations noted earlier suggest that LF phonons originate from the surface/interface of Si NC in $SiO_2$. Molecular dynamics simulation shows that oxidation leads to stretching of Si-Si bonds around the interface only[53] due to presence of different suboxides at the interface.[54-57] Further, vibrational density of states calculations show that at the surface of nanocrystals, phonon frequencies up to 494 cm$^{-1}$ can be observed.[58] Consistent with this, we attribute LF phonons to stretching vibration of Si-Si bonds at the Si-$SiO_2$ interface. The phonon frequency variation in the range from 495 to 510 cm$^{-1}$ can be attributed to the variation in local $SiO_x$ environment seen by Si-Si bonds at the surface of the Si NC. This will also lead to a Lorentzian line shape of Si-Si stretching vibration with appropriate width depending on $SiO_x$ environment as observed by us.

The blue shift in LF phonons (Fig. 2(d)) accompanied with betterment in crystalline order during LHC experiment can be understood in terms of more crystalline Si-Si bonding environment leading to more ordered Si-$SiO_2$ interface. This will be further discussed in the next section.

The attribution of HF phonons observed in the range 515 - 519 cm$^{-1}$ to the core phonon of Si NCs is consistent with observation that no change in frequency, width & line shape is observed during the LHC experiment.

*5.1.    Variation of Si phonons originating at the interface (LF phonons) during the LHC experiment ::*

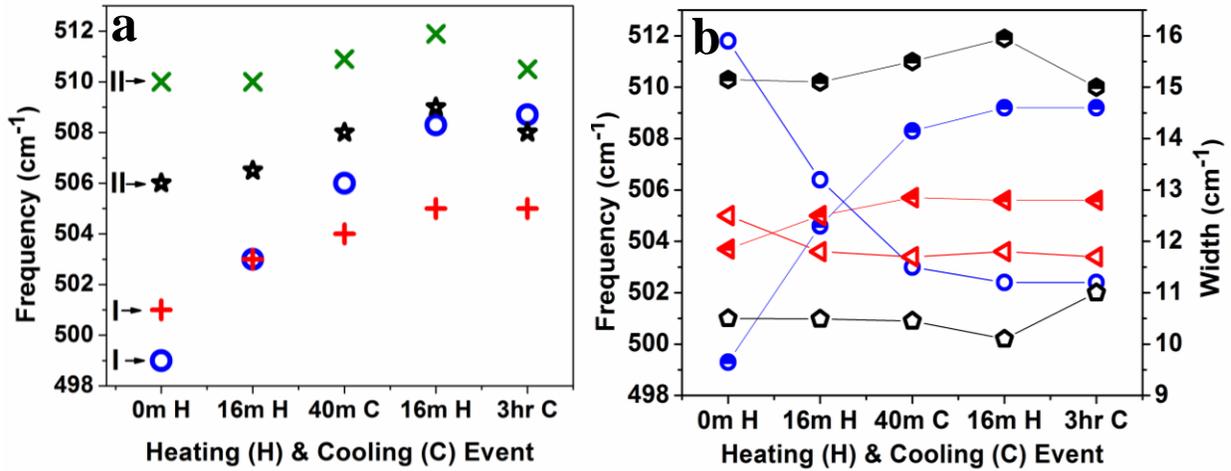

Fig. 3. a)  Change in phonon frequency during LHC experiment for LF phonons in the range ~ 495 to 510 cm⁻¹  showing two patterns : I and II and b) Change in phonon frequency and corresponding FWHM during LHC experiment for LF phonons. Half filled symbols show frequency and open symbols show corresponding FWHM.

Variation in frequency of various LF phonons in the range 495 - 510 cm⁻¹ is plotted in Fig. 3(a) during 5 steps of LHC. We have performed the same experiment at many such sites in the said frequency range in all the samples. It is interesting to note that primarily there are two types of patterns in the change of frequencies during the LHC experiment, i) in the first pattern (type I), initially blue shift is observed and leading to freezing of phonon frequency at the end of the experiment and ii) in the second pattern (type II), freezing of phonon frequency is observed in initial steps and thereafter, blue shift on heating and redshift on cooling. There is another interesting point to note here that the variation of phonon frequency and FWHM are mirror images of each other (see Fig. 3(b)) during the LHC experiment. For lower phonon frequency, FWHM is higher and one can also see that blue shift in phonon frequency is always accompanied with betterment in crystalline order of the interface for all the frequencies plotted here. This behavior of the phonon frequency and corresponding FWHM is similar as due to change in the temperature of a crystalline material. For bulk and nanocrystalline Si, temperature dependent Raman studies show that red shift of the optical phonon mode by $\sim$ 10 cm$^{-1}$ and $\sim$ 15 cm$^{-1}$ respectively, for temperature rise from 300 to 800 K.[59, 60] Thus to know the local temperature during the LHC experiment, we have performed Stokes/anti-Stokes Raman measurements for several LF and HF phonons as well as on the sites where both are observed simultaneously. Local temperature calculated using intensity ratio of Stokes to anti-Stokes $(I_S/I_{AS})$[59] of HF phonon is found to be the room temperature (300 K). This is consistent with the fact that no change in HF phonon is observed. However, local temperature calculated using LF phonon (interface frequencies) is found to be in the range of $\sim$ 550 - 700 K (at spatially different positions). Moreover, continuous blue shift doesn't show a continuous decrease in temperature (Fig. 4) indicating that there does not exist any correlation between frequency shift of the phonon and the temperature obtained from Stoke/anti-Stoke Raman measurements. It is important to note that the two different local temperatures at a site, obtained using two different phonons cannot be explained by conventional understanding. Further, after 16 mins (after 2$^{nd}$ step in Fig. 4) of heating, we observe a decrease in measured temperature, suggesting that there are inconsistencies in these results. The plausible explanation for this is that the local temperature calculated from $I_S/I_{AS}$ data using interface/surface phonon (LF phonons) may be incorrect. This can happen, if excitation source is close to one of the real energy levels of the system, leading to resonance Raman scattering. In this case the Raman cross-section is dominated by the resonance term and hence the $I_S/I_{AS}$ ratio is no more proportional to phonon occupation number. Therefore, the $I_S/I_{AS}$ ratio can give incorrect value and hence cannot be used for calculation of the local temperature. Resonance Raman scattering is indeed observed for LF phonons, whereas HF phonons show non-resonant behavior and detailed study of the same is discussed separately.[52] This further supports the understanding that LF and HF phonons have different origin.

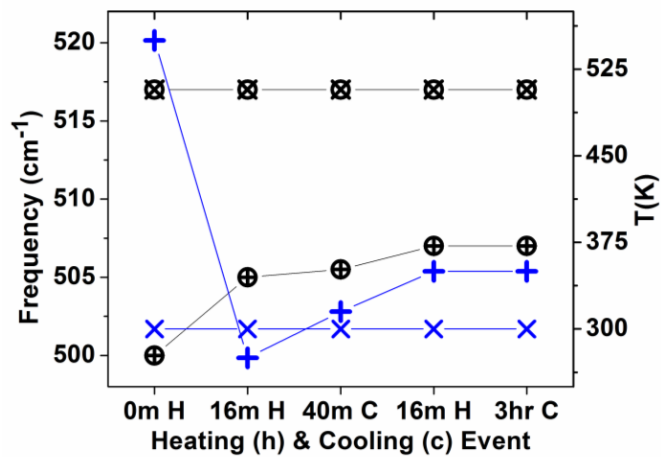

Fig. 4. Change in frequency and calculated local temperature during LHC experiment for LF and HF phonons. Symbols (+) and (×) in circle show LF and HF phonon and symbols (+) and (×) show corresponding local temperature.

Study of surface/interface in such dynamic conditions (laser irradiation) along with Raman spectroscopy is not feasible using any conventional experimental technique such as X-ray photoelectron spectroscopy[57] and Auger electron spectroscopy[61]. Further, earlier molecular dynamics calculation suggested strong contribution of surface vibrational density of states in Si-SiO$_2$.[58, 62] Therefore, to gain further insight into surface / interface phonons we carry out *ab initio* DFT and TDDFT based calculations of Raman spectra of optimized Si cluster (Si$_{41}$) terminated by Oxygen (O) and Hydrogen (H) atoms as discussed in the next subsection.

5.2.   *Calculation of Raman spectra of Si cluster (Si$_{41}$) with different terminations:*

In order to support our experimental results as well as to gain microscopic understanding, we perform density functional theory[63] based electronic structure calculations for a medium size silicon cluster ($Si_{41}$), with bulk-like tetrahedral coordination, using Amsterdam Density Functional (ADF) code[64]. We consider two possible terminations for silicon atoms present at the surface of the cluster, namely (a) oxygen and (b) hydrogen atoms, in order to avoid the dangling bonds. The former termination is used to mimic the Si cluster/$SiO_2$ interface. The calculations of ground state properties have been carried out by using triple-$\xi$ Slater-type orbital (STO) basis set with two added polarization functions (TZ2P basis set of ADF basis set library)[65] along with generalized gradient approximation (given by Perdew-Burke-Ernzerhof (PBE)[65]) exchange-correlation (XC) functional. The geometric structures of the cluster with the two above mentioned terminations, and with bulk-like initial positions, are fully relaxed using the BFGS technique until the norm of the energy gradient and energy attain values less than $10^{-4}$ au and $10^{-6}$ au respectively.

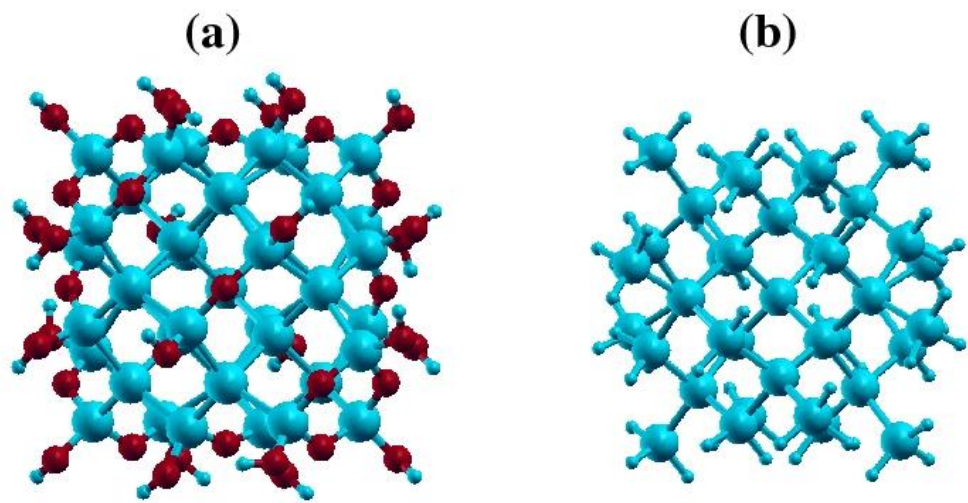

**(a)**                    **(b)**

Fig. 5. Ball and stick model for the optimized geometric structures of $Si_{41}$ cluster with (a) oxygen ($Si_{41}O_{42}H_{24}$) and (b) hydrogen atoms ($Si_{41}H_{60}$) terminations which are obtained by DFT based electronic structure calculation with PBE XC functional and TZ2P basis set. The balls with colors cyan (large), red (medium) and cyan (small) represent Si, O and H atoms respectively.

The fully optimized geometric structures of $Si_{41}$ cluster with (a) oxygen ($Si_{41}O_{42}H_{24}$) and (b) hydrogen atoms ($Si_{41}H_{60}$) terminations are shown in Fig. 5. The results of our detailed geometric analysis are summarized in Table 1. We observe from this table that the mean values for Si-Si bond lengths (2.38 in $Si_{41}O_{42}H_{24}$ and 2.37 Å in $Si_{41}H_{60}$) are slightly higher than its corresponding value of 2.34 Å in bulk silicon. This indicates that the interaction between Si atoms in the cluster is little weaker than that in the bulk form. The maximum variation in the value of bond length, with respect to its bulk value is 0.07 Å. The trend of increase in the mean values for these angles is given in 4th - 6th columns of Table 1. The standard deviations (SD) for both the bond lengths and bond angles are also given in Table 1. The calculated mean values (109.23 and 107.64° for $Si_{41}O_{42}H_{24}$ and $Si_{41}H_{60}$ respectively) of bond angles in the core region indicate that the local environment of Si atom in this region is bulk-like since these values are close to the corresponding bond angle (109.47°) between the Si atoms in bulk. However, we observe large modifications in the mean values of the bond angles between Si atoms as we go to the surface region. Our results show that there exist two different types of bond angles in the surface region: one value is much lower and another being higher than that of in the core region. In case of $Si_{41}O_{42}H_{24}$, we also find different types of bond angles between Si and O atoms in the surface region. Overall, our results of geometric analysis clearly indicate that the hybridization between Si atoms in the core region is nearly $sp^3$-like and it is modified in to a mixture of $sp^2$ and $sp^3$ in the surface regions.

| System | Bond length (Å) | | Bond angle (º) | | | | | |
|--------|------|------|---------------|------|------|------|------|------|
| **$Si_{41}O_{42}H_{24}$** | **Si-Si** | **Si-O** | **Si-Si-Si (Core)** | **Si-Si-Si (Core/Surf.)** | **Si-Si-Si (Core/Surf.)** | **O-Si-O** | **Si-O-Si** | **Si-O-Si** |
| Mean | 2.38 | 1.66 | 109.23 | 100.23 | 117.10 | 107.79 | 126.32 | 147.18 |
| SD | 0.016 | 0.009 | 1.05 | 1.53 | 2.87 | 2.07 | 0.46 | 0.50 |
| **$Si_{41}H_{60}$** | **Si-Si** | **Si-H** | **Si-Si-Si (Core)** | **Si-Si-Si (Core/Surf.)** | **Si-Si-Si (Core/Surf.)** | **H-Si-H** | **-** | **-** |
| Mean | 2.37 | 1.504 | 107.64 | 103.06 | 117.40 | 107.78 | - | - |
| SD | 0.016 | 0.005 | 1.07 | 1.15 | 0.11 | 0.60 | - | - |

**Table 1**: The results for the bond lengths and bond angles of $Si_{41}$ cluster with (a) oxygen ($Si_{41}O_{42}H_{24}$) and (b) hydrogen atoms ($Si_{41}H_{60}$) terminations obtained by DFT based electronic structure calculation with PBE XC functional and TZ2P basis set.

In order to gain more insight into the experimental results for Raman spectra discussed above we employ these optimized structures to perform calculations of Raman spectra. For this purpose we make use of linear response theory within TDDFT which is implemented in RESPONSE module of the ADF code.[64] The intensity of Raman spectra is calculated by estimating the first derivative of the polarizability with respect to normal coordinates. We use the same TZ2P basis set for the calculations of polarizability. It is well known that for performing calculations of response properties within TDDFT, it is required to use approximate forms for the XC potential at two different levels. At the first level, we use standard PBE XC potential for the calculation of ground-state Kohn-Sham orbitals and their energies. The second level of approximation is needed for XC kernel which determines the XC contribution to the screening of applied fields and for this we use the reasonably accurate adiabatic local density approximation (ALDA). The calculated Raman spectra of $Si_{41}$ cluster with (a) oxygen ($Si_{41}O_{42}H_{24}$) and (b) hydrogen atoms ($Si_{41}H_{60}$) terminations are presented in Fig. 6. It is observed from Fig. 6 that there is a strong peak at 512 cm$^{-1}$ in Raman spectra of oxygen terminated Si$_{41}$ cluster. Similarly, we also observe a strong peak at 499 cm$^{-1}$ in Raman spectra of hydrogen terminated Si$_{41}$ cluster. It is important to note that the frequencies of these two strong peaks observed in Si$_{41}$ cluster with oxygen and hydrogen atoms terminations are different and they are much lower than the corresponding Raman peak ~ 521 cm$^{-1}$ in bulk silicon. Furthermore, our detailed analysis shows that the normal modes of vibrations corresponding to these strong peaks contain motions of Si atoms lie on the surface of Si$_{41}$ cluster, whereas, high frequency modes with lower intensity are involving all atoms. Thus, our calculated vibrational frequency (~ 512 cm$^{-1}$) of the strong peak in oxygen terminated cluster can be considered to be analogous to the observed phonon frequencies (495 - 510 cm$^{-1}$) which are originating from the surface of Si NCs in the presence of SiO$_2$ matrix. Calculation of Raman spectrum further shows that surface phonon contribution is dominant in small cluster. This further supports the interpretation that LF phonons originate from surface/interface of smaller Si NCs, whereas HF phonons originate from core of larger Si NCs.

On the other hand, the calculated frequency of a strong peak in Raman spectra of hydrogen terminated cluster is much lower (~ 499 cm$^{-1}$) than the corresponding frequencies in oxygen terminated cluster. These results are also consistent with the experimental data reported in the literature. In hydrogen passivated Si NCs (NCs Si: H), Raman spectra show phonon frequencies in the range 500 - 517 cm$^{-1}$ with 514.5 nm as excitation laser.[66-68] Whereas, Raman spectrum of NCs Si : H shows phonon frequencies ~ 439, 462, 478, 486, 504 cm$^{-1}$, lower than the observed in Si-SiO$_2$ nanocomposites with 496.5 nm as excitation laser.[69] This is consistent with results of our calculations. Further, it is relevant to note here that observance of surface/interface phonons (LF) is laser excitation dependent.[52] It is evident from our calculations that Si atoms at the surface of cluster and their local environments play an important role in determining Raman spectra. This is not surprising, as it is well established that for smaller size NCs, the surface optical phonon becomes observable. With further decrease in size of NCs, surface phonon becomes stronger as compared to core phonon (confined phonon) i.e. confinement effect due to size becomes ineffective.[70] Surface optical phonon is lower in frequency and it's frequency depends on the surrounding medium.[70]

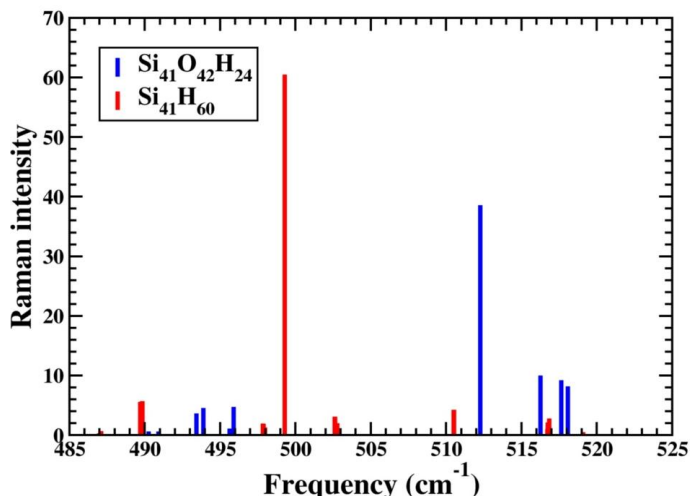

Fig. 6. Raman spectra of $Si_{41}$ cluster with (a) oxygen ($Si_{41}O_{42}H_{24}$) and (b) hydrogen ($Si_{41}H_{60}$) terminations obtained by employing DFT/TDDFT based response property calculation.

*5.3 Understanding of the behavior of LF phonons during LHC experiment:*

Variation found in bond angle and bond length (Table 1) due to $sp^3$ and $sp^2$ hybridization at the surface can well be correlated to large width for Raman spectra of LF phonons, originating at the interface, which sees this variation. Our results suggest that this low laser power density is enough to make rearrangements at the interface of Si NCs. However, no measurable changes could be observed in the Raman spectra of larger size NCs. This can be resultant of two effects, first that smaller NC may have higher temperature rise at same laser PD due to porosity at the interface compared to that of larger NCs due to different nature of their interface. Molecular dynamics calculations by Soulairol et al. show that for smaller NCs the interface is porous i.e. incoherent and as the size increases, it's porosity decreases and interface tends to become more coherent for larger NCs embedded in $SiO_2$ matrix.[62] Secondly, although total surface area is larger for larger NCs, surface to volume ratio is much smaller for larger NCs for this surface/interface effect to be observable. Since, local temperature measurement from LF phonons is not feasible as noted above, therefore, we have performed power dependent Raman measurements on the site where LF and HF phonons are observed to get further evidence that the observed effect of temperature. When, laser PD is increased from 10 kW/cm$^2$ to 300 kW/cm$^2$, it is observed that LF phonon red shifts by ~ 1-2 cm$^{-1}$ initially, whereas HF phonon shows no significant change (red shift ~ 0.2 - 0.4 cm$^{-1}$) (Fig. 7). Continuous heating for half an hour at this power density further shows blue shift in LF phonons and no change in HF phonons. Initial red shift observed in here for LF phonons may not be due to rise in temperature as it does not continue while laser is ON. Instead this observed red shift of 1-2 cm$^{-1}$ in LF phonons may be due to relaxation of the compressive stress, predicted for small NCs in $SiO_2$ matrix.[62] Next, the blue shift observed in LF phonons during the LHC experiment as well as with continuous laser heating can be correlated to either generation of the compressive stress or correction of Si-O bond to Si-Si bond at the surface/interface. However, betterment in crystalline order during blue shift is suggestive of increase in coherency of interface during laser heating/annealing. The heat/energy provided during LHC experiment can lead to the conversion of Si-Ox-Si bonding to Si-Si bonding and in process filling of voids i.e. increases in the coherency of the interface and also leading to blue shift of Si phonon. Theoretical investigation by Jiang et al. also shows that temperature recycling improves the sharpness of the interface.[71] This can also lead to freezing of value of LF phonons in some cases as observed during the LHC experiment (Type 1 : Fig. 3a), as once interface is coherent, small energy provided may not be sufficient to change the environment due to presence of matrix. However, when the interface is more coherent, this energy is then transferred to the matrix and compressive stress will be generated in Si NCs [20], leading to a small blue shift in LF phonon frequency (~ 1 - 2 cm$^{-1}$). In this case, therefore, cooling, results in relaxation of this stress, leading to small red shift (1-2 cm$^{-1}$, Type-II: Fig 3(a)). This observed red shift is only for the higher LF phonons (frequencies ~ 506 - 510 cm$^{-1}$) during cooling process. It is important to observe that lower LF phonons (frequencies ~ 495 -505 cm$^{-1}$) show larger blue shift (8-10 cm$^{-1}$) on heating and freezing is observed for these phonons only at the end of LHC experiment, whereas, higher frequency LF phonons show freezing at the initial steps of LHC. This is consistent with the interpretation of LF phonon originating at the interface and lowering of the Si phonon frequency (leading to LF phonons) is due to higher levels of defects at the interface i.e. variation in the surrounding environment of Si NCs, which gets corrected leading to the blue shift. This in turn depends on size to a large extent as nature of bonding at the interface changes with size of a nanocrystal.[57] For Si-SiO$_2$ nanocomposite, it has been predicted theoretically that Si=O bond forms at the interface leading to the formation of Si$^{3+}$ suboxides, as the size of Si NCs decreases.[54] This is also later confirmed experimentally using XPS technique by Kim et al.[57] Increase in size of Si NCs leads to increase in formation of Si-O bonds leading to higher intensities of Si$^{2+}$ suboxides as compared to Si$^{3+}$ suboxides.[57] This is  further investigated using XPS measurements as described in the following.

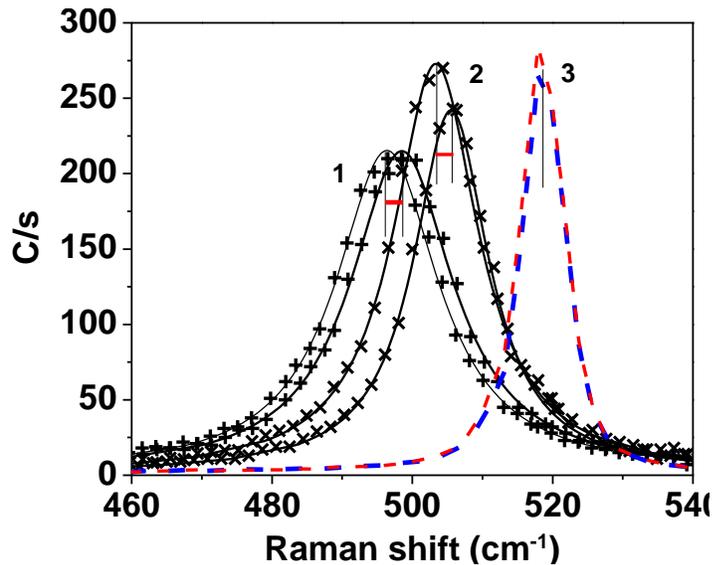

Figure 7. Change in LF (1, 2) and HF phonons (3) at laser PD ~ 10 kW/cm$^2$ and 300 kW/cm$^2$. Raw data is shown by symbol and Lorentzian fit to the data by solid line for LF phonons.

Two samples, grown under similar conditions but at different times, wherein laser energy may be different are studied using Raman mapping. The i) 15 mutilayers (E1) and ii) 10 mutilayers (E7) Si-SiO$_2$ are grown, which shows abundant presence of LF phonons and very rare presence of LF phonons, respectively. To see if this difference reflects in XPS data as per our interpretation, XPS is performed on E1 and E7 (Fig 8). Since, The beam size in XPS is ~ 1 cm * 1 cm and thus covering the full sample as noted in the experimental section. For this reason, average signal expected from XPS measurement is corroborated with the statistical information obtained from the Raman data. The XPS data is fitted using Gaussian-Lorentzian line shapes for the Si 2$p$ core-level profile after linear background subtraction using XPS peakfit software.[72] By deconvoluting the spectrum, three chemical structures corresponding to Si$^{4+}$ (B. E. ~ 103.4 eV), Si$^{3+}$ (B. E. ~ 102.3 eV) and Si$^{2+}$ (B.E. ~ 101.2 eV) are observed. XPS gives relative contribution of $Si^{4+}$ with that of $Si^{3+}$ and $Si^{2+}$ in both the samples. Relative intensities (I) of $Si^{3+}$ and $Si^{2+}$ are found to be different in both the samples. $I(Si^{3+})/I(Si^{2+})$ = 2.6 and 0.6 for samples E1 and E7 respectively. As noted above, the $I(Si^{3+})/I(Si^{2+})$ is expected to be > 1 where smaller nanocrystals are more in number and $I(Si^{3+})/I(Si^{2+})$ is expected to be < 1 where larger nanocrystals are more in number. According to our interpretation, E1 should have smaller size NCs and hence should have higher $I(Si^{3+})/I(Si^{2+})$ ratio compared to that of E7. The suboxide peaks corresponding to $Si^{3+}$ and $Si^{2+}$ observed in our sample for E1 and E7 is suggestive of the same i.e. higher number of smaller size NCs in sample E1 as compared to that of sample E7. This independently confirms our attribution of LF phonons as surface/interface phonons of Si NCs in $SiO_2$ matrix and that variation in the Si phonon frequency as due to local variation of Si-$SiO_2$ interface for LF phonons. It is important to note that the other samples (numbered as E2-E6) which shows similar occurrence of LF phonons, also show XPS spectra similar to sample E1.

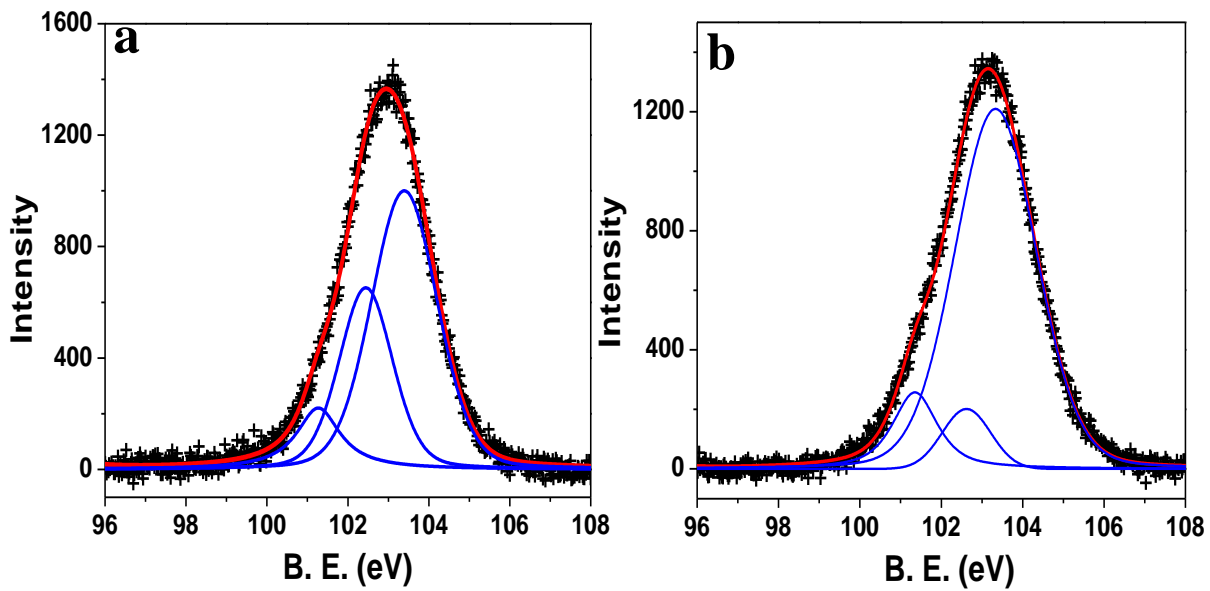

Fig. 8: XPS spectra of sample a) E1 and b) E7 showing difference in relative ratio of $Si^{3+}$ and $Si^{2+}$.

## 6. Conclusion

We have studied PLD grown Si-SiO$_2$ multilayer nanocomposites with different Si layer thickness leading to different size Si nanocrystals. Spatial variation of Si phonon frequencies in the range 495 - 519 cm$^{-1}$ has been investigated using Raman spectroscopy. To understand the origin of Si phonon frequencies in low frequency region, Raman spectroscopy monitored local laser heating/annealing and cooling experiment is devised on the desired frequency. On the basis of line shape variation and behavior of Si-phonon during LHC the Si-phonons frequencies in the range 495 - 510 cm$^{-1}$ (LF phonon : Lorentzian line shape) and 515 - 519 cm$^{-1}$ (HF phonons : asymmetric line shape) are attributed to the interface of Si NCs and SiO$_2$ matrix and core of Si NCs respectively. The important point to note is that blue shift of LF phonons during LHC is accompanied with the betterment in crystalline order taking it closer to crystalline Si phonon frequency; however the frequency freezes at frequencies lower than HF phonons due to SiO$_2$ matrix. Stokes/anti-Stokes measurements suggest that LF phonons are observable due to Resonance enhancement of Raman signal. This understanding is also consistent with the observation of such a strong signal coming from interface of Si-SiO$_2$. Furthermore, our results of DFT/ TDDFT based Raman spectra calculations for Si$_{41}$ cluster with oxygen and hydrogen terminations establish the dominant contribution of interface phonon i.e. surface phonon with different terminations for smaller size Si nanocrystals. In summary, LF phonons originate from the smaller size NCs, where phonon frequency is governed by dominant surface/interface effect and HF phonon originates from larger size Si NCs, where core phonon is dominant and it's frequency is governed by the confinement effect. XPS measurements of two similar samples grown at different times, showing large contrast in number of LF phonon sites can be well correlated with the contrast in formation of various suboxides, supporting the correspondence between occurrence of  LF phonons and presence of smaller size Si nanocrystals with extended interface in Si-SiO$_2$ nanocomposite. The intermediate frequencies are further investigated in this light and will be published elsewhere.[52]

**Acknowledgment:**

It is a pleasure to acknowledge Dr. H. S. Rawat and Dr. G. S. Lodha for the continual support to this work.  Ms Ekta Rani wishes to acknowledge help of Mr. Viresh Nayak in Matlab programming, Mr. A. D. Wadekar for performing XPS measurements and Homi Bhabha National Institute, India for providing research fellowship during the course of this work. Authors also thank the scientific computing group of computer centre, RRCAT for the support in running the ADF code.